\documentclass{ws-p8-50x6-00}

\begin{document}

\title{Hyperonic Crystallization in Hadronic Matter}

\author{M.A. P\'erez-Garc{\'i}a$^1$}

\address{E-mail: aperez{\char64}pinon.ccu.uniovi.es}

\author{J. D{\'i}az-Alonso$^{1,2}$, L. Mornas$^{1}$, 
J.P. Su\'{a}rez$^{1}$}

\address{$(^1)$ Dpto. de  Fisica, Universidad de Oviedo. 
Avda. Calvo Sotelo 18, E-33007 Oviedo, Asturias, Spain 
\\$(^2)$ DARC, Observatoire de Paris - Meudon, 
F-92195 Meudon, France}  

\maketitle

\abstracts{
The possible formation of a spatially ordered phase in neutron star 
matter is investigated in a model where hyperonic impurities are
localized on the nodes of a cubic lattice.}

We propose a method for analyzing the formation of spatially ordered
configurations in hadronic matter, related to the crystallization of the
hyperonic sector in a lattice. The method allows the determination of 
the density ranges where such ordered configurations are energetically 
favoured with respect to the usual gaseous configurations. This leads 
to the determination of the parameters of a first order phase transition
from the fluid to the crystallized state as well as the equation of 
state of the plasma in the ordered phase. We show some preliminary 
results obtained by the application of the method to a very simplified 
model which shows the efficiency of the mechanism of confinement of
 the hyperons in an ordered phase proposed here. The eventual existence 
of such a new phase should induce changes in the structure and cooling 
of dense stars. 

At densities about 1.5 times nuclear saturation and beyond, hyperons
are present in hadronic matter. The equation of state taking into 
account the presence of nucleons and hyperons which interact through 
the exchange of several mesons has been extensively studied using 
phenomenological lagrangian models \cite{lattimer}.
In the mean field approximation, the hyperonic component (as the other 
hadronic components in the ground state) is supposed to form a 
spatially uniformly distributed Fermi fluid undergoing the action of 
the mesonic mean field created by the whole baryonic distribution 
in the plasma.

The hadronic interaction in our model is reduced to the exchange of 
scalar and vector meson fields. Such a description is the simplest one 
allowing for an acceptable fit of nuclear saturation \cite{wal}. We 
consider a unique species of neutral hyperons. The interaction 
Lagrangian for every baryon species $B$ reads
\begin{equation}
L_{I}=\sum\limits_{B=p,n,h}g_{\sigma B}\overline{\Psi }\sigma \Psi
+g_{\omega B}\overline{\Psi }\gamma ^{\mu }\omega _{\mu }\Psi
\end{equation}
where the coupling constants to the meson fields are weaker for 
hyperons than for nucleons. 
The leptonic sector is reduced for simplicity to the electron. 

Using relativistic condensed matter techniques, we investigate the 
possibility of the existence of energetically favourable solid 
configurations, where pairs of antiparallel spin hyperons are confined 
on the nodes of a regular lattice in the ground state of the hadronic 
plasma. In this situation the hyperons should behave as impurities 
which induce a redistribution of the surrounding nucleons. 
Nevertheless, as a first approximation to the solution for the dynamics 
of the nucleon in the medium, we consider the lattice of neutral 
hyperons as a uniform background, in an analogous way to the ''free 
electron model'' in a Coulomb lattice \cite{ashcroft}. 

In evaluating the energetic contribution of a crystal array of hyperons
surrounded by the nucleon liquid component in the plasma, the hyperons 
are assumed to be localized on the nodes of a cubic lattice under the 
action of their mutual interaction and the mean fields created by the 
surrounding nucleons. In this way the total potential at every lattice 
site is approximated by a harmonic potential which is determined 
selfconsistently as the superposition of the potentials created by 
the gaussian clouds of two antiparallel spin hyperons on all the other 
nodes of the lattice, to which the interaction with the mean field of 
the nucleon sector is added. In a first approximation we neglect the 
central and spin-spin interaction between the hyperons in the same 
node. We shall consider such interaction in future work, but we have 
already verified that their effect improves the confining character 
of the whole potential and facilitates the crystallization. 
In calculating the lattice potential we have also introduced monopolar 
form factors accounting for the composite structure of the hyperon. 
In this analysis, vanishing temperature is assumed. 

For the nucleonic background in the crystal, a relativistic Hartree
approximation is used to calculate the contribution to the total energy
density of the system, as in the case of the liquid ground state to 
which it is compared. 

The relative abundances of protons, neutrons and hyperons 
can be now obtained by establishing the equations of 
$\beta$-equilibrium between all these particles. The chemical 
potentials of protons, neutrons and electrons are the Fermi energies 
of these particles which are in Fermi gaseous phases. The chemical 
potential of the hyperons are calculated as the energy gained by the 
system when a new hyperon is added in the $ n = 1 $ level of the 
harmonic potential in a node. 
We need also the equations for the mean scalar and vector fields 
generated by the uniform background of nucleons. The abundance of 
hyperons resulting from this system is an ingredient in the calculation 
of the potential created by the lattice on every node.

Consequently, for the periodic configuration the self-consistent set of
equations for the beta equilibrium of chemical potentials in the plasma
coupled to the mean field equations, and the self consistent equations 
for the confining fields in the nodes are 
\begin{eqnarray}
n_{electrons} &=& n_{protons}\qquad \mbox{\rm (charge neutrality)} \\
\mu_{neutrons} &=& \mu_{protons}+\mu _{electrons}  \\
\mu _{neutrons} &=& \mu _{hyperons} \\
n_{baryons} &=& n_{neutrons}+n_{protons}+n_{hyperons} \\
m_{\sigma }^{2}<\sigma +\sigma _{ext}> &=& 
g_{\sigma N}<\overline{\Psi }\Psi >+g_{\sigma H} n_{hyperons} \\
m_{\omega }^{2}<\omega ^{0}+\omega _{ext}^{0}>
&=& -g_{\omega N}<\overline{\Psi }\gamma ^{0}\Psi >
-g_{\omega H} n_{hyperons} \\
m_{\sigma }^{2}<\sigma _{ext}> &=& g_{\sigma H} n_{hyperons} \\
m_{\omega }^{2}<\omega _{ext}^{0}> &=& -g_{\omega H} n_{hyperons} 
\end{eqnarray}
The crystal size cell $a$ is related to the hyperonic density
$n_{hyperons}$ of hyperons paired in a S-state. 
\begin{equation}
n_{hyperons}=\frac{2}{a^{3}}
\end{equation}
The solution of this system gives the magnitudes characterizing the
configuration, such as crystal cell size, characteristic oscillation 
frequency and width of hyperon wave functions, the energy levels, etc 
as functions of the total baryonic density \cite{yo}. 

In Figure 1 we show the confining potential in a node of the hyperonic 
lattice for a set of values of the parameters of the model 
($a$=2.2142 fm, cut-off for the scalar and vector fields, 
$\Lambda_{\sigma} = 1.4$ GeV, $\Lambda_{\omega}=2.5$ GeV, value of the 
coupling constants $g_{\sigma H}/g_{\sigma N}=0.6$, 
$g_{\omega H}/g_{\omega N}=0.65$ taking the Walecka values for the 
meson-nucleon interaction). This figure has been obtained by the self 
consistent solution of the equations relating the confining potential 
to the width of the hyperonic wave function. 

Figure 2 shows the cell size in the lattice as a function of baryonic 
density obtained by asuming the solid configuration to be stable at 
every baryonic density. 

It is worth mentioning that the screening effects of the 
nucleonic plasma have to be treated properly in order to consider 
the modification from the screened HH interaction to the vacuum one, 
as well as the mixing effects of the hadronic fields. This work is 
actually in progress. 

To determine whether the transition from a spatially uniform hyperonic
liquid phase to an ordered one takes place inside the plasma, and the 
range of thermodynamic variables where the crystal phase is 
energetically favoured, the free energies of both phases have to be 
compared. The densities of both phases in the transition region is 
determined by looking for the cross point between the liquid and 
solid configuration lines in the diagram of partial hyperon pressure 
versus hyperonic chemical potential.

This phase transition is of first order, leading to a change in the
equation of state of the plasma with a modification of the mechanical
properties which should have consequences on the hydrostatic 
equilibrium of neutron stars and modify the analysis of the cooling 
processes.

\begin{figure}[t]
\begin{minipage}{12cm}   
\mbox{%
\parbox{5.5cm}{%
\epsfxsize=13.5pc 
\epsfbox{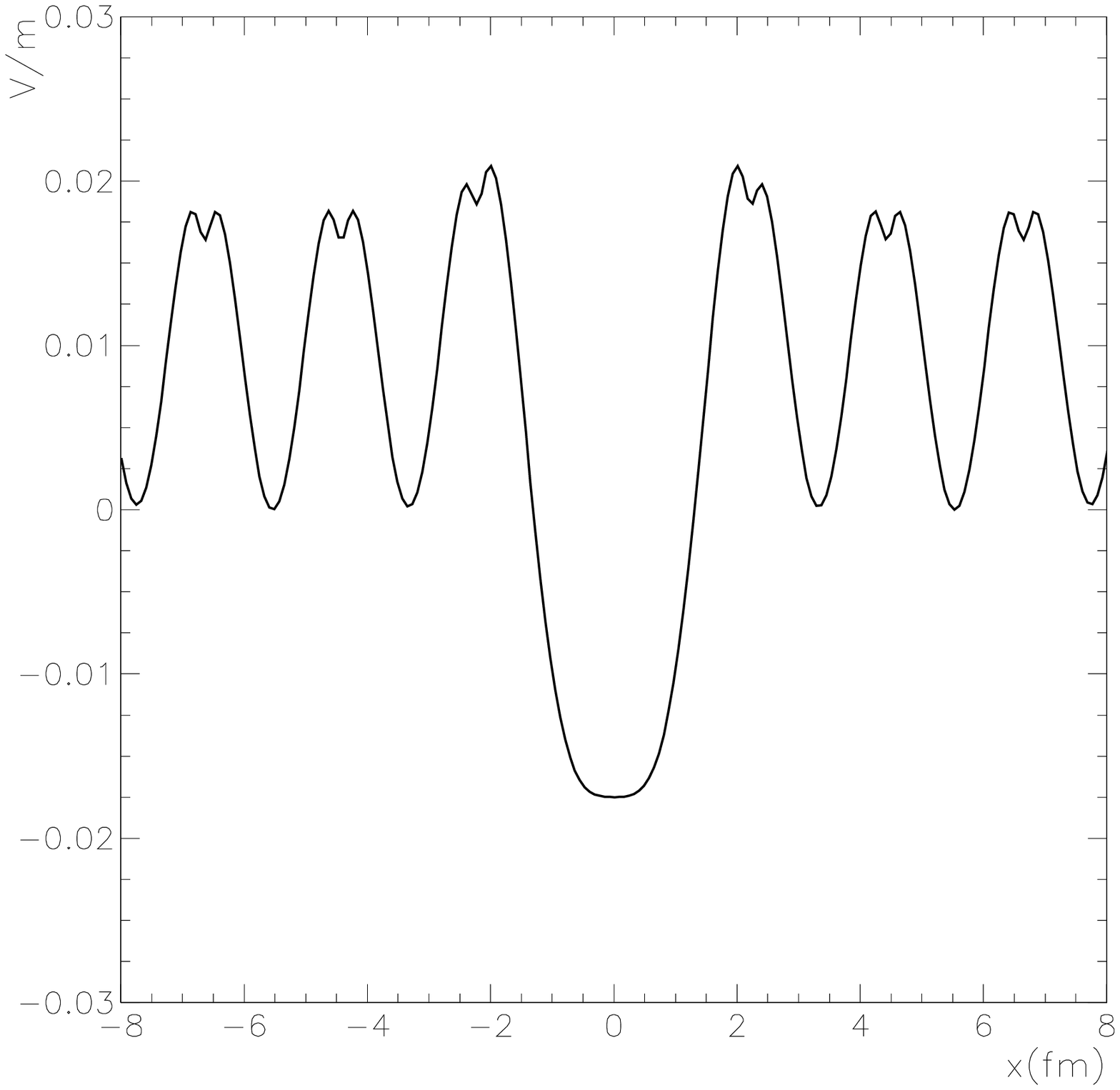}
\vskip -1cm 
\caption{Self consistent confining potential}
}
\parbox{0.5cm}{\phantom{aaa}}
\parbox{5.5cm}{%
\epsfxsize=13.5pc 
\epsfbox{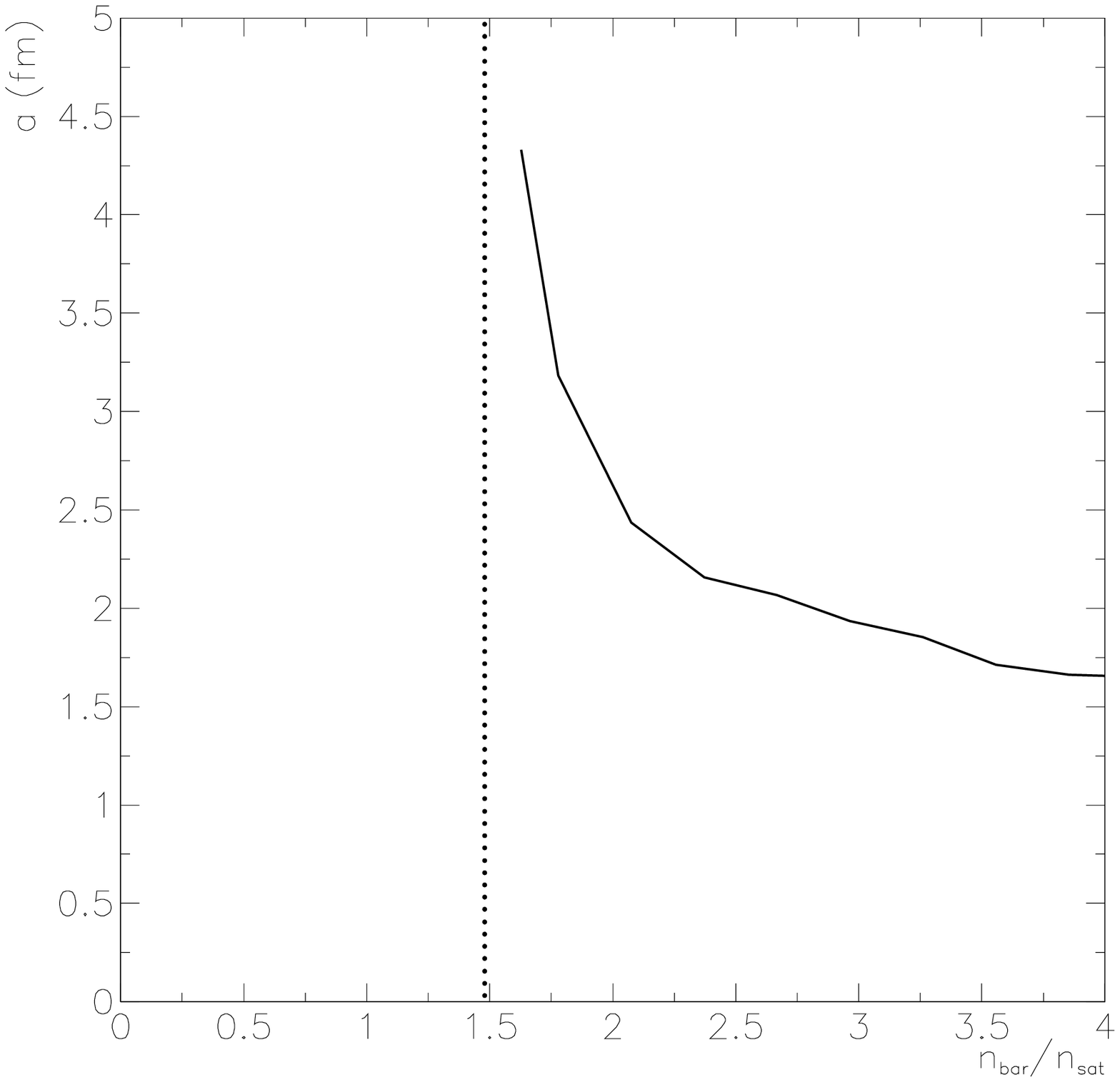} 
\vskip -1cm 
\caption{Lattice size as a function of density}
}
}
\end{minipage} 
\end{figure}

\end{document}